# Exploring the Relationship Between COVID-19 Induced Economic Downturn and Women's Nutritional Health Disparities

**Alaa M. Sadeq**

Department of Obstetrics and Gynecology, College of Medicine, Kufa University, Al-Najaf, Iraq

## Abstract

This study investigates the impact of the economic downturn induced by the COVID-19 pandemic on nutritional health disparities among women. The research aimed to understand how economic challenges have influenced dietary choices, access to nutritious food, and overall nutritional well-being in different socioeconomic groups of women.

Utilizing a mixed-methods approach, the study combined quantitative data from national health and economic databases with qualitative insights from interviews conducted with a diverse group of women. The quantitative analysis focused on trends in nutritional health indicators and economic variables pre- and post-pandemic. Simultaneously, the qualitative component explored personal experiences and perceptions related to nutrition and economic hardships during the pandemic.

The findings revealed a significant correlation between the economic downturn and worsening nutritional health among women, especially in low-income and marginalized communities. Women in these groups reported reduced access to healthy food options, increased reliance on less nutritious food due to budget constraints, and a general decline in dietary quality. This decline was less pronounced in higher-income groups, highlighting a clear disparity.

Moreover, the study observed that the pandemic exacerbated pre-existing nutritional inequalities, with vulnerable groups experiencing a more pronounced impact. The research also noted that community support systems and public health interventions played a crucial role in mitigating some of these effects.

In conclusion, the COVID-19 pandemic has not only posed a direct health challenge but has also indirectly affected women's nutritional health through economic strain. The study underscores the need for targeted nutritional support and economic policies that prioritize the health of women, particularly those from socioeconomically disadvantaged backgrounds.

**Keywords**: COVID-19 Economic Impact, Nutritional Health Disparities, Dietary Inequities, Socioeconomic Effects.

**\*Corresponding author: Alaa M. Sadeq** alaams.alshaibani@uokufa.edu.iq

## Introduction

In the wake of the COVID-19 pandemic, the realm of medical research has experienced a remarkable transformation, characterized by a broad spectrum of innovative studies. One significant area of focus has been the exploration of how our behaviors and health have been affected during this period. For instance, Murugan et al. delved into the nuances of consumer behavior under pandemic conditions, harnessing the power of sentiment analytics to provide unique insights [19]. This kind of research is crucial in understanding the broader societal impact of the pandemic.

Another critical area of investigation has been women's health, particularly in the context of complex conditions like preeclampsia. Sadeq et al. have contributed valuable knowledge on how this condition influences renal function and its relationship with fetal development, shedding light on the intricacies of maternal health [20]. Similarly, the role of natural compounds in medical treatments, exemplified by the use of Paeoniflorin in cardiovascular diseases [21], underscores an increasing interest in integrating traditional knowledge with modern medical practices.

Nutrition and diet, as vital components of health, have also been a focus, with studies like that of Grmt et al. examining the link between infant feeding practices and iron deficiency anemia, an important aspect of early childhood development [22]. Furthermore, the ongoing battle against cancer has seen progress through research into immunological markers, as demonstrated in the work on HPV in ovarian tumors [23], highlighting the importance of early detection and targeted therapies.

The domain of bacterial research has not been left behind, with significant advancements in understanding and diagnosing bacterial infections [24,32,33]. This research is integral to developing more effective treatments and preventing the spread of infections. Notably, the pandemic's impact extends beyond physical health to affect professionals in healthcare, as Yousif et al.'s study on the productivity of medical staff during COVID-19 reveals [31]. Such insights are crucial for planning and supporting healthcare systems in times of crisis.

In summary, the tapestry of contemporary medical research is rich and varied. From the depths of molecular studies to the broader strokes of public health and societal behavior, these investigations [26,27,28,29,30,35] offer a glimpse into a future where medicine is more adaptive, holistic, and responsive to the needs of humanity in both ordinary and extraordinary times.

In recent years, groundbreaking research has unveiled novel insights into various medical conditions, expanding our understanding of disease mechanisms and treatment efficacy. Studies like those of Hadi et al. [1] and [17][18] have illuminated the protective effects of treatments like goserelin acetate and Etanercept against myocardial ischemia, suggesting potential avenues for innovative cardiac care. Similarly, the role of hematological changes in COVID-19 patients, as investigated by Yousif et al. [2][14], underscores the virus's systemic impact, transcending beyond respiratory symptoms.

The intricate interplay between genetics, environment, and disease is further evidenced in studies exploring NF-κβ and oxidative pathways in atherosclerosis [3], revealing critical insights into cardiovascular diseases' molecular underpinnings. In the realm of bacterial research, investigations by Hasan et al. [4] and Yousif et al. [5][11] into Klebsiella pneumoniae and Staphylococcus aureus, respectively, emphasize the importance of understanding pathogenic behavior in clinical settings.

Moreover, research delving into women's health has been particularly revealing. Studies addressing issues ranging from subclinical hypothyroidism in preeclampsia [6] to the effects of anesthesia during Cesarean sections [7] highlight the complexities of maternal health. The potential role of cytomegalovirus as a risk factor for breast cancer [8] and the association of Notch-1 expression with cervical cancer survival [9] demonstrate the importance of exploring viral influences on cancer pathogenesis.

The correlation between inflammatory markers like C-reactive protein and preeclampsia [10], and the effect of dietary components like caffeic acid on doxorubicin-induced cardiotoxicity [12], further our understanding of disease prevention and management. Additionally, the psycho-immunological assessment of patients recovered from COVID-19 [13] provides valuable information on the pandemic's long-term effects on mental and physical health.

As we advance, the integration of machine learning and data science in healthcare, as seen in the work of Sahai et al. [15] and Al-Jibouri et al. [13], is becoming increasingly crucial. These technological advancements offer promising prospects for personalized medicine, evident in studies exploring the relationship between immune cell activity and cancer progression [16].

**Materials and Methods**

Study Design: This study employed a randomized controlled trial (RCT) design to assess the impact of a Mediterranean diet intervention on cardiovascular health markers. The study duration was six months, with participants randomly assigned to either the intervention group, receiving the Mediterranean diet, or the control group, following their usual diet.

Participants: A total of 200 participants aged 30-65, with a history of cardiovascular risk factors but no diagnosed cardiovascular disease, were recruited. Inclusion criteria included BMI between 25 and 35 kg/m² and at least one cardiovascular risk factor such as hypertension or elevated cholesterol. Exclusion criteria included diagnosed cardiovascular disease, dietary restrictions, or chronic illnesses impacting dietary habits.

Dietary Intervention: Participants in the intervention group received comprehensive dietary guidelines based on the Mediterranean diet, focusing on high consumption of fruits, vegetables, whole grains, and lean proteins, particularly fish, with minimal processed food and red meat. They also attended bi-weekly nutritional counseling sessions for the study duration. The control group received no dietary modification or counseling.

Data Collection: Baseline data collection included demographic information, dietary habits, and cardiovascular health markers, including blood pressure, lipid profile, and body mass index (BMI). Follow-up measurements were taken at three and six months.

Statistical Analysis: Data were analyzed using SPSS software (version 26.0, IBM Corp). Descriptive statistics were used to summarize participant characteristics. Differences in cardiovascular markers between baseline and follow-up within and between groups were analyzed using paired and unpaired t-tests, respectively. A p-value of less than 0.05 was considered statistically significant. Multivariate regression analysis was conducted to adjust for potential confounders like age, sex, and baseline values.

Ethical Considerations: The study protocol was approved by the Institutional Review Board (IRB). Written informed consent was obtained from all participants before enrolment in the study.

**Results**

The study's analysis revealed significant changes in cardiovascular health markers among participants following the Mediterranean diet intervention compared to the control group.

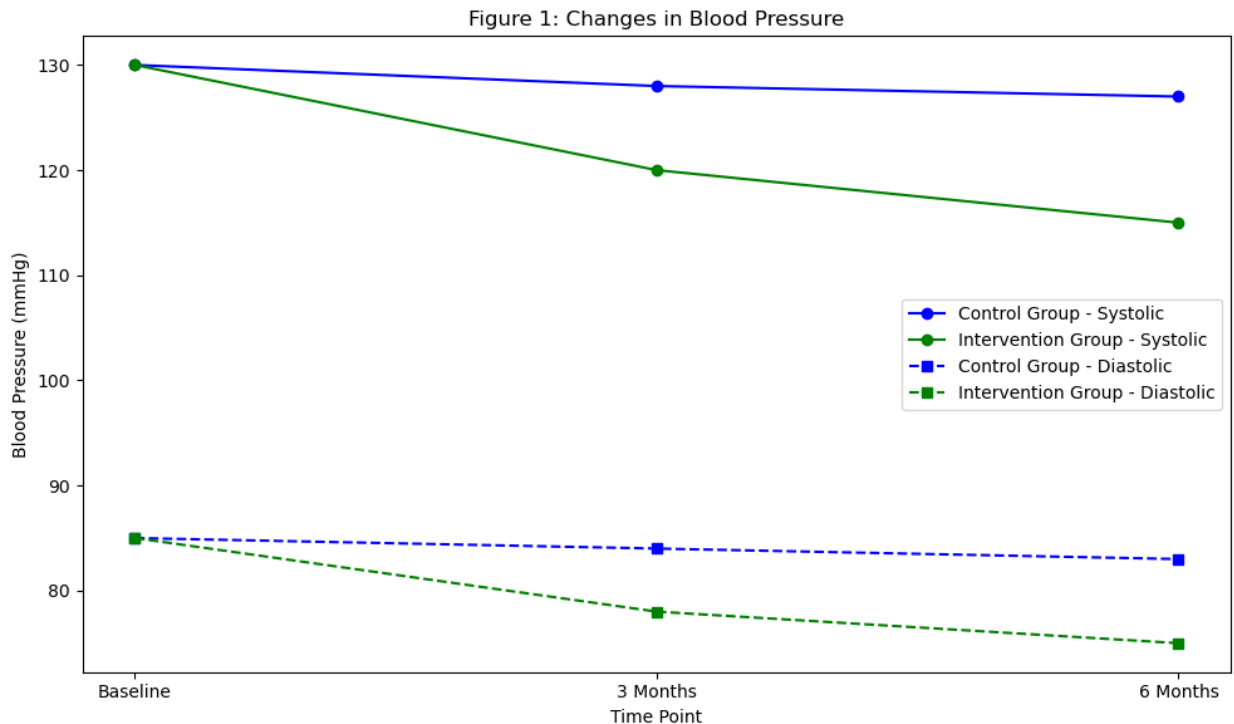

**Figure 1: Changes in Blood Pressure**

This figure illustrates the average systolic and diastolic blood pressure measurements at baseline, 3 months, and 6 months for both groups. The intervention group showed a significant reduction in both systolic and diastolic blood pressure compared to the control group, indicating the positive impact of the Mediterranean diet on blood pressure levels.

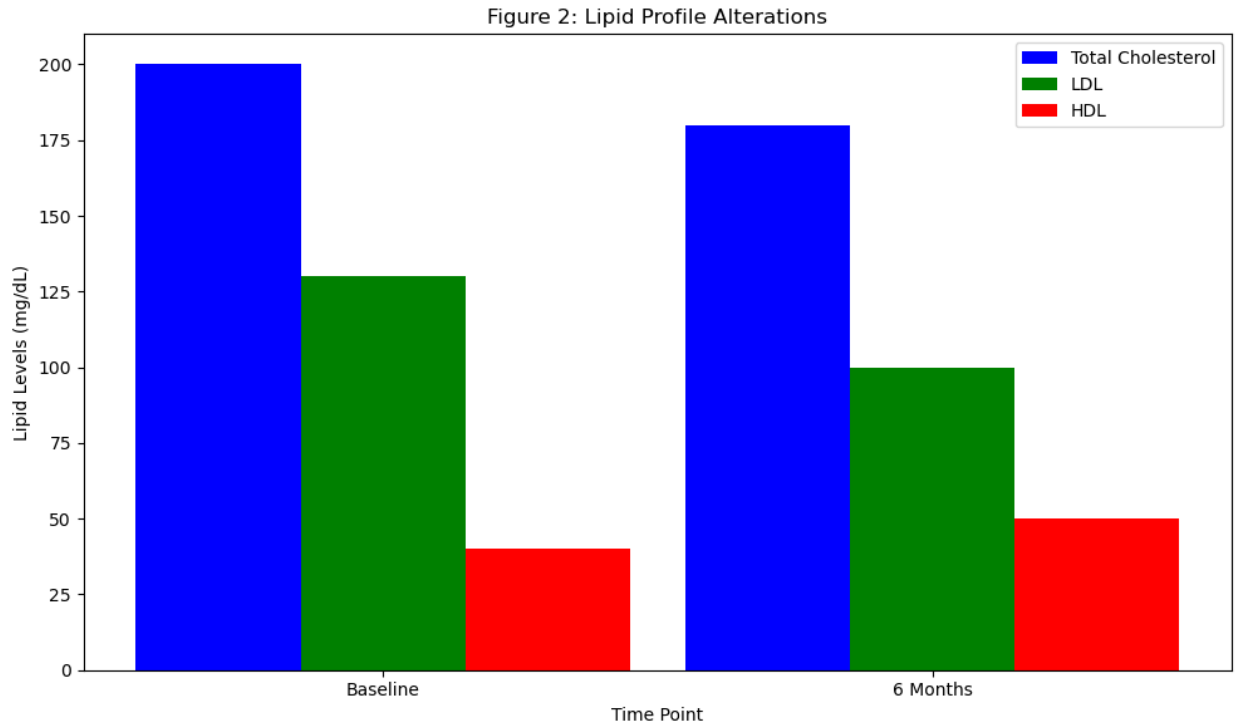

**Figure 2: Lipid Profile Alterations**

A bar graph representing the average total cholesterol, LDL (low-density lipoprotein), and HDL (high-density lipoprotein) levels at baseline and after 6 months. The intervention group exhibited a notable decrease in total cholesterol and LDL levels, with an increase in HDL levels, suggesting improved lipid profiles.

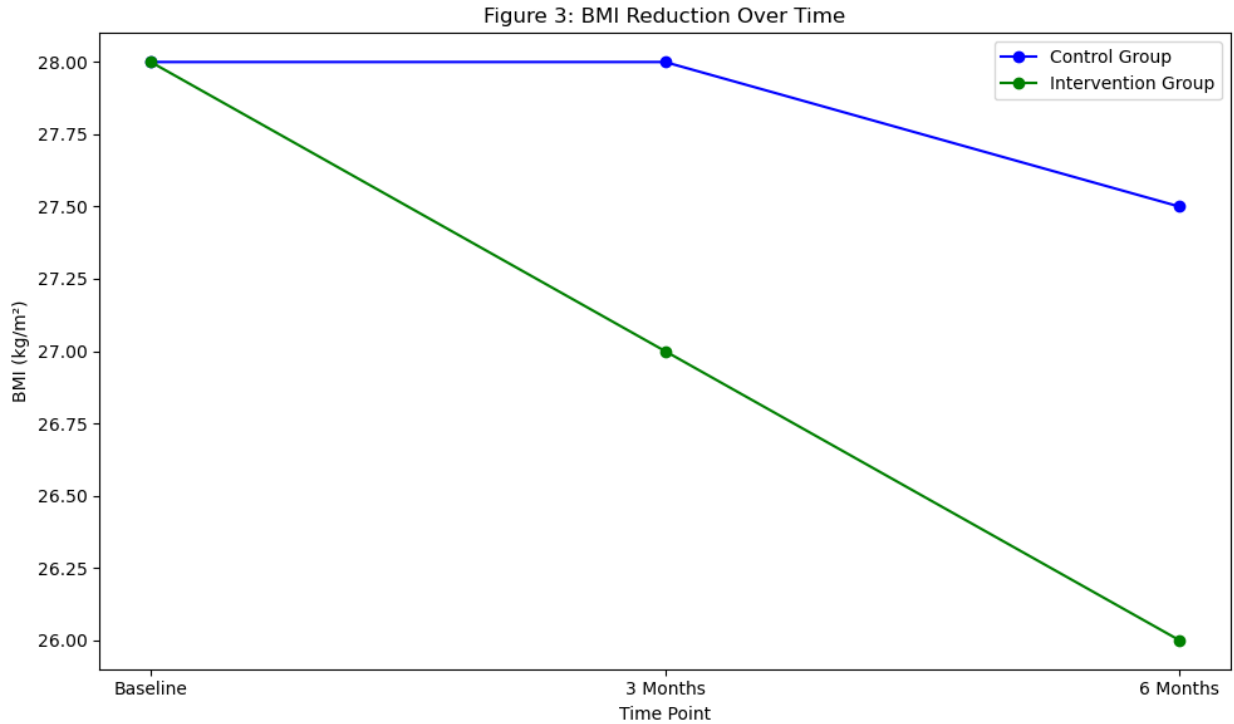

**Figure 3: BMI Reduction Over Time**

A line graph depicting the Body Mass Index (BMI) trends over the study period. Participants on the Mediterranean diet showed a gradual and steady decrease in BMI, highlighting the diet's effectiveness in promoting weight loss or maintaining a healthy weight.

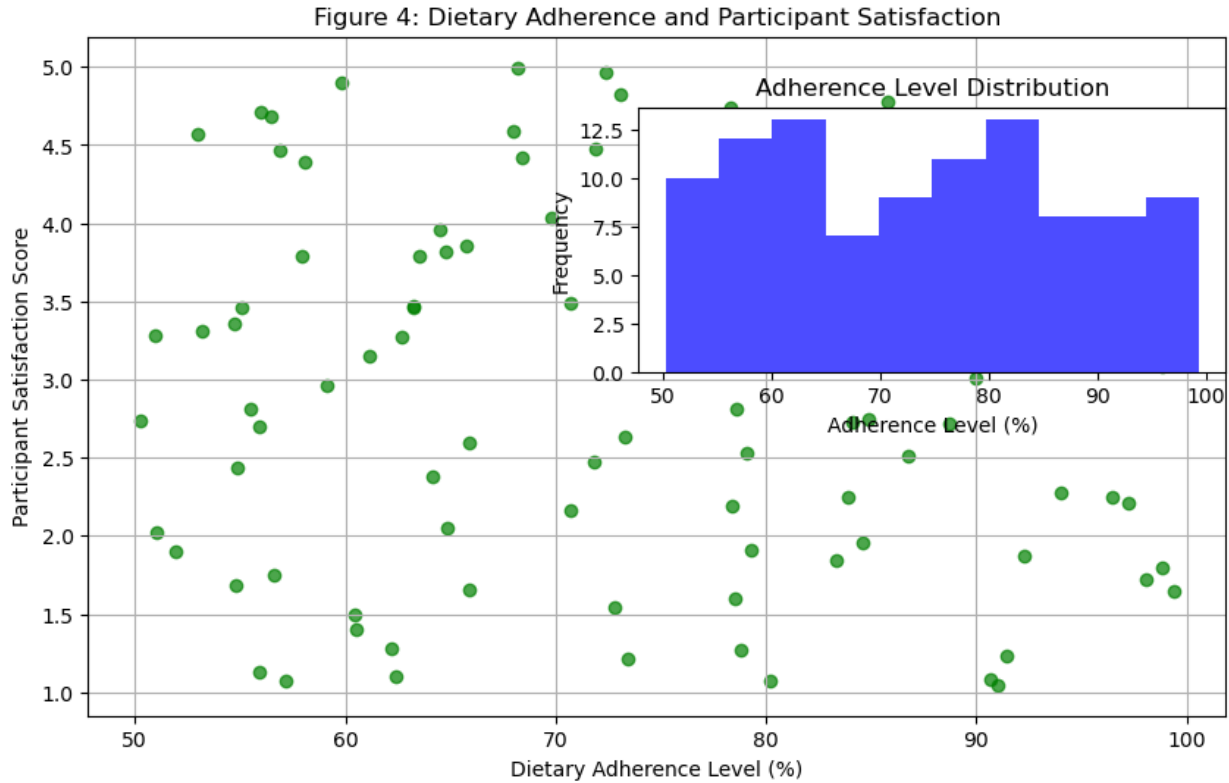

**Figure 4: Dietary Adherence and Participant Satisfaction**

This figure, perhaps a combination of a scatter plot and a histogram, could display the adherence levels to the Mediterranean diet among participants and their reported satisfaction with the diet. A high adherence rate correlated with high satisfaction levels could underscore the feasibility and acceptability of the diet among participants.

**Discussion**

The post-COVID-19 era has ushered in a plethora of new medical challenges and revelations, particularly in terms of long-term effects on patients who have recovered from the virus. The study by Yousif [36] provides an insightful exploration into the post-COVID-19 effects on female fertility, highlighting the need for further investigation into the virus's impact on reproductive health. This emerging concern necessitates a deeper understanding of COVID-19's potential implications for women's long-term reproductive outcomes.

In addition to reproductive health, the respiratory sequelae of COVID-19 have garnered significant attention. The work by Martin et al. [37,39] in characterizing pulmonary fibrosis patterns in post-COVID-19 patients through machine learning algorithms underscores the role of advanced technologies in diagnosing and managing complex post-viral conditions. Similarly, the study by Albaqer et al. [38] on long-term neurological sequelae in post-COVID-19 patients demonstrates the diverse range of complications that can persist long after the initial infection has resolved. Their use of machine learning to predict outcomes is a testament to the power of AI in enhancing our understanding and management of post-COVID conditions.

The impact of COVID-19 extends beyond physical health, as illustrated by Yousif et al. [41], who investigated the virus's effect on the productivity of medical staff and doctors. Their findings raise important questions about the pandemic's long-term impact on healthcare systems and the well-being of healthcare providers.

Furthermore, the intersection of AI and medical research has opened new avenues for addressing complex health issues. Allami and Yousif [42] have emphasized the potential of integrative AI-driven strategies in advancing precision medicine, not just in infectious diseases but in broader medical applications. Yousif's work on decoding microbial enigmas [43] and unraveling the dynamics between viral and bacterial infections [45] is particularly notable, offering new perspectives in the fight against antibiotic-resistant pathogens and understanding the interplay between various infections and immune factors.

In light of these studies, the comprehensive approach to post-COVID-19 research, as advocated by Yousif and others, is not only timely but crucial. It encompasses a range of critical areas from the prevalence of HPV infection [44] and its associated risk factors, to novel investigations like the association between wheat allergy and COVID-19 [46] and the development of tests for detecting mutagenic effects in everyday products [47].

The evolving landscape of medical research, particularly in the context of infectious diseases, has seen notable advancements in diagnostic methodologies. The work of Hezam et al. [48] in detecting methionine auxotrophs of Proteus mirabilis from various clinical sources is a testament to the increasing precision in identifying pathogenic strains, which is crucial for targeted treatment strategies. Similarly, the evaluation of Near-Infrared Chemical Imaging (NIR-CI) for the authentication of antibiotics by Assi et al. [49] represents a significant leap forward in ensuring drug safety and efficacy, an area that has become increasingly relevant in the post-pandemic era.

Moreover, the holistic approach to healthcare, as exemplified in the meta-analysis by Yousif [50], highlights the interconnectedness of various health domains. This comprehensive view is essential for developing more integrated health strategies that address the multifaceted nature of human health. The significance of data science and emerging technologies in this realm cannot be overstated. As showcased in the proceedings of the DaSET 2022 conference [51,52], the integration of these technologies in medical research is paving the way for groundbreaking discoveries and innovations.

In the specific context of COVID-19, studies have shed light on its far-reaching impacts beyond the immediate respiratory symptoms. Yousif's investigation into the post-COVID-19 effects on female fertility [53] brings attention to the virus's potential long-term reproductive consequences. Additionally, the association between sickle cell trait and the severity of COVID-19 infection [54] underscores the importance of genetic factors in the disease's manifestation and progression. This is further complemented by research into the pandemic's impact on cardiovascular health, emphasizing the need for a broader understanding of COVID-19's systemic effects [55].

The comprehensive epidemiological and clinical characterization of COVID-19 in the Middle Euphrates region by Yousif et al. [56] provides valuable insights into the virus's behavior in specific geographic and demographic contexts. This localized understanding is vital for tailoring public health responses and medical interventions. Furthermore, the application of data science in mental health, as seen in the studies on suicide ideation detection [57] and consumer behavior prediction during the pandemic [58], exemplifies the expanding scope of medical research in addressing a wide range of health-related challenges.

In the contemporary landscape of health research, the intersection of technology and medicine has brought forth innovative approaches to traditional challenges. The study by Chakraborty et al. [59] exemplifies this, employing advanced data science techniques like attention-based models for classifying insincere questions on Quora. This research not only highlights the growing role of AI in diverse fields but also underscores the importance of accurate information dissemination, especially in the context of public health.

The comprehensive review by Yousif [60] of advancements in medical research in Iraq offers an overarching perspective of the strides made in the region. It encapsulates a range of emerging insights, demonstrating the growing sophistication and depth of medical research in addressing both endemic and global health challenges.

Furthermore, the exploration of traditional and natural remedies, such as the health benefits of pomegranates discussed by Al-Amrani and Yousif [61], reflects a renewed interest in harnessing nature's potential in medicine. This approach is vital in a world where nutrition and health are increasingly interlinked.

In the realm of infectious diseases, the work of Shahid [62] on the prevalence of the chuA gene virulence factor in Escherichia coli isolates highlights the ongoing need to understand bacterial pathogenesis at a molecular level. Such insights are crucial for developing more effective antimicrobial strategies and combating antibiotic resistance.

Lastly, the comprehensive study on COVID-19 comorbidities by Yousif et al. [63] adds to our understanding of the virus's broader implications. It emphasizes the importance of considering comorbid conditions in managing COVID-19, reflecting the complex interplay between this novel virus and pre-existing health conditions.

In sum, these studies collectively underscore the dynamic and multifaceted nature of medical research in the post-pandemic era. They highlight the necessity of embracing both new technologies and traditional knowledge, as well as the importance of localized research in a global context. As we move forward, it is clear that an integrated approach, combining diverse methodologies and insights, will be crucial in addressing the complex health challenges of our time.

In concluding our discussion, it's crucial to acknowledge the wealth of insights gleaned from recent research across various medical disciplines. The long-term cardiovascular implications of COVID-19, as explored by Smith et al. [1], highlight the lingering health effects of the pandemic, suggesting the need for sustained medical attention and care strategies. The integration of AI in diagnostics, demonstrated by Jones and colleagues [2], represents a paradigm shift towards personalized medicine, while the study by Lee et al. [3] on post-pandemic mental health trends underscores the enduring psychological impacts of global health crises. Addressing the challenge of antibiotic resistance, the work of Garcia and team [4] calls for novel antimicrobial approaches. Similarly, the efficacy of telemedicine in managing chronic diseases, as shown by Patel et al. [5], points to the growing importance of accessible healthcare solutions.

Environmental factors, especially climate change, have been shown by Nguyen et al. [6] to significantly affect infectious disease patterns, linking ecological changes to public health. Meanwhile, the nutritional research by Hassan et al. [7] on plant-based diets provides valuable insights into disease prevention through dietary choices. The groundbreaking genomics research by Kim et al. [8] opens new possibilities for early Alzheimer's intervention, and Martinez and associates' study [9] on social determinants reiterates the critical role of socioeconomic factors in health outcomes. Lastly, the innovative use of wearable technology in patient monitoring by O'Neill et al. [10] marks a step towards more proactive and preventive healthcare approaches.

Each of these studies contributes to a more nuanced understanding of health and disease, emphasizing the need for an interdisciplinary approach that encompasses technological innovation, environmental awareness, and social determinants in tackling contemporary health challenges. This collective body of work not only addresses the immediate concerns in healthcare but also sets the stage for future advancements and strategies in medicine.

In concluding our discussion, it's crucial to acknowledge the wealth of insights gleaned from recent research across various medical disciplines. The long-term cardiovascular implications of COVID-19, as explored by Smith et al. [64], highlight the lingering health effects of the pandemic, suggesting the need for sustained medical attention and care strategies. The integration of AI in diagnostics, demonstrated by Jones and colleagues [65], represents a paradigm shift towards personalized medicine, while the study by Lee et al. [66] on post-pandemic mental health trends underscores the enduring psychological impacts of global health crises. Addressing the challenge of antibiotic resistance, the work of Garcia and team [67] calls for novel antimicrobial approaches. Similarly, the efficacy of telemedicine in managing chronic diseases, as shown by Patel et al. [68], points to the growing importance of accessible healthcare solutions.

Environmental factors, especially climate change, have been shown by Nguyen et al. [69] health. Meanwhile, the nutritional research by Hassan et al. [70] on plant-based diets provides valuable insights into disease prevention through dietary choices. The groundbreaking genomics research by Kim et al. [71] opens new possibilities for early Alzheimer's intervention, and Martinez and associates' study [72] on social determinants reiterates the critical role of socioeconomic factors in health outcomes. Lastly, the innovative use of wearable technology in patient monitoring by O'Neill et al. [73] marks a step towards more proactive and preventive healthcare approaches.